# Efficient Feasibility Analysis for Real-Time Systems with EDF scheduling*

Karsten Albers, Frank Slomka
Department of Computer Science
University of Oldenburg
Ammerländer Heerstraße 114-118
26111 Oldenburg, Germany
{albers, slomka}@informatik.uni-oldenburg.de

**Abstract**

*This paper presents new fast exact feasibility tests for uniprocessor real-time systems using preemptive EDF scheduling. Task sets which are accepted by previously described sufficient tests will be evaluated in nearly the same time as with the old tests by the new algorithms. Many task sets are not accepted by the earlier tests despite them beeing feasible. These task sets will be evaluated by the new algorithms a lot faster than with known exact feasibility tests. Therefore it is possible to use them for many applications for which only sufficient test are suitable. Additionally this paper shows that the best previous known sufficient test, the best known feasibility bound and the best known approximation algorithm can be derived from these new tests. In result this leads to an integrated schedulability theory for EDF.*

## 1. Introduction

The analysis of the time behavior of embedded real-time systems is essential for the automation of the design process. A formal verification which guarantees all deadlines in a real-time system would be the best. This verification is called feasibility test. Three different kinds of tests are available:
- Exact tests with long execution times or simple models [2], [3], [11].
- Fast sufficient tests which fails to accept feasible task sets, especially those with high utilizations [9], [12].
- Approximations, which are allowing an adjustment of performance and acceptance rate [1], [8].

For many applications an exact test or an approximation with a high acceptance rate must be used. For many task sets a fast sufficient test is adequate.

This paper proposes two new tests for preemptive *EDF* scheduling, which have a performance comparable with the sufficient tests for those tasks sets which can be recognized by these sufficient tests and outperforms the existing exact tests by orders of magnitude. These algorithms are tested and compared using a large amount of randomly generated task sets and some examples from literature.

In Section 2 the analysis model is introduced, Section 3 gives an introduction to the existing feasibility tests and

* The research described is supported by the Deutsche Forschungsgemeinschaft under grant SL 47/1-1

the concepts needed further in this paper. It contains the first contribution of this work: the proof of equivalence between the sufficient test given by Devi [9] and the superposition approximation described in [1]. It is shown that the test by Devi is a special case of the superposition approach. Section 4 presents as the main contribution of the work the new fast sufficient and necessary feasibility tests and also the proof for deriving the feasibility bound out of these tests. In Section 5 the new test algorithms are evaluated followed by a conclusion in Section 6.

## 2. Model

We consider the sporadic task system consisting of a set of tasks $\Gamma=\{\tau_1,...,\tau_n\}$. Each task $\tau_i$ is described by
- an initial release time (or phase) $\phi_i$,
- a relative deadline $D_i$ (measured from the release time),
- a worst-case execution time $C_i$ and
- a minimal distance (or period) $T_i$ between two instances of a task.

An invocation of a task is called a job, and the $k^{th}$ invocation of each job is denoted $\tau_{i,k}$. In the following only the synchronous case is considered, so the first jobs of all tasks are released simultaneously. This is a common assumption which also leads to a sufficient test for the asynchronous case [14]. A good sufficient condition for the asynchronous case is proposed in [13]. It is based on the tests improved in this paper.

The specific utilization of a task is the part of the capacity which is needed for executing this task ($U(\tau) = C_\tau/T_\tau$). The utilization $U$ of the system is the sum of the specific utilizations of all tasks. We consider the uniprocessor feasibility test. The scheduling is done using earliest deadline first (*EDF*) which is known to be optimal [12].

To keep the explanations in this paper simple the sporadic task model is used. The new tests can be extended to more advanced task models. Especially the extension for the event stream model [11] is easy by following the definitions proposed in [1].

## 3. Background and related work

The feasibility test for uniprocessor systems using EDF with deadlines smaller then the periods of the tasks is known to be Co-NP-hard [3]. There are only feasibility tests with pseudo-polynomial complexity available so far. It is unknown whether the problem has polynomial complexity. In the following an introduction to the relevant feasibility tests of the literature is given. An overview is given in [14].

### 3.1. Sufficient tests

For a restricted version of the given model with the condition $T_i = D_i$ for each task $\tau_i$, Liu and Layland [12] proved that the task system is feasible if the utilization $U \leq 1$.

### 3.2. Test by Devi

Recently Devi [9] proposed a test which allows task systems with deadlines smaller than periods with a reduced complexity:

**Def. 1: Test Devi** [9]**:** *A task system $\Gamma$, arranged in order of non-decreasing relative deadlines is feasible using EDF scheduling if $n = |\Gamma|$ and*

$$\forall (1 \leq k \leq n) \left| \left( \sum_{i=1}^{k} \frac{C_i}{T_i} + \frac{1}{D_k} \cdot \sum_{i=1}^{k} \left( \frac{T_i - min(T_i, D_i)}{T_i} \right) \cdot C_i \leq 1 \right) \right.$$

Unfortunately this test is only sufficient. It allows a fast evaluation and acceptance of task sets in many cases.

### 3.3. Processor demand test

In [3] Baruah et al. proposed an exact test for task systems with deadlines shorter than their periods. The main idea was to calculate the maximum demand of all tasks within a time interval and compare it with the available capacity of the processor. The demand is the cumulated worst-case execution time of the relevant jobs. The maximum demand for an interval $I$ can be found if all tasks are released synchronous at the beginning of $I$ and only those jobs are considered which have both their release time and their absolute deadline within $I$.

**Def. 2: Demand bound function $Dbf(I)$** [2]**:** *The maximum cumulated execution requirement of jobs having both request time and deadline within interval $I$*

$$dbf(I, \Gamma) = \sum_{\forall \tau_i \in \Gamma \wedge I \geq D_i} \left\lfloor \frac{I - D_i}{T_i} + 1 \right\rfloor \cdot C_i$$

It is possible to split the demand bound function into demand bound functions for each single task. $dbf(I,\tau)$ is the cumulated execution requirement of only one task. The available capacity of execution time in an interval $I$ is exactly the length of the interval. Testing the demand bound function for all intervals against the capacity leads to an exact test. To make the test tractable it is possible to calculate a maximum test interval ($I_{max}$) which is an upper bound (feasibility bound) and it is only necessary to test the demand bound function for intervals lower than $I_{max}$.

**Def. 3: Processor demand test** [2]**:** *A task system $\Gamma$ is feasible if and only if $dbf(I,\Gamma) \leq I$ for all $I_{max} = (U/(1-U))\{max(T_i - D_i)\}$.*

It is only necessary to check those intervals where the value of $dbf(I,\Gamma)$ is changed. The absolute deadlines of all jobs determine these intervals. Other values for $I_{max}$ are discussed in Section 4.3.

The test has a pseudo-polynomial complexity if the considered utilization is bound by a constant. The problem is that the runtime of the algorithm depends not only on the utilization but also on the ratio of the different periods and deadlines in the task set [1]. If the task set contains tasks with small periods and tasks with large periods, the runtime can become quite large. See the experiments given in Section 5 for more details.

### 3.4. Approximation by superposition

Recently, several approximations have been proposed to solve the complexity problem of the demand bound approach: One by Chakraborty et al. [8] and the superposition approach [1]. They bridge the gap between the fast but only sufficient test of Devi and the necessary but slow processor demand test. The algorithms are only sufficient, but the degree of sufficiency can be adjusted.

The main idea of the superposition approach is to test only a selectable limited amount of test intervals for each task and use an approximation to compensate the remaining test intervals.

**Def. 4: Approximated task demand bound function $dbf'(I,\tau)$** [1]**:** *An upper bound for the task demand bound function considering only the jobs up to the selectable maximum test interval $Im(\tau)$ exactly.*

$$dbf'(I, \tau) = \begin{cases} dbf(Im(\tau), \tau) + \frac{C_\tau}{T_\tau} \cdot (I - Im(\tau)) & I > Im(\tau) \\ dbf(I, \tau) & I \leq Im(\tau) \end{cases}$$

The approximation is using the specific utilization of the task. In the special case of considering only the first job of each task, the maximum test interval is equal to the deadline of the task. The approximated demand bound function is a superposition of the approximated task demand bound functions.

**Def. 5: Approximated demand bound function [1] $dbf'(I,\Gamma)$:**

$$dbf'(I, \Gamma) = \sum_{\forall \tau \in \Gamma} dbf'(I, \tau)$$

One advantage of this approximation is that it is not necessary to calculate $dbf'(I,\Gamma)$ for each job separately.

**Lemma 1: Super position test:** *A task system $\Gamma$ is feasible if $dbf'(I,\Gamma) \leq I$ for all $I \leq I_{max}$*

**Def. 6: SuperPos(x) :** *The super position test which calculates at the maximum the first x jobs of each task exactly (test level x) is called SuperPos(x)*

A higher level of the superpostion test leads to a higher acceptance rate but also to a longer execution time of the test. Figure 1 shows measured acceptance rates using different SuperPos(x). The superposition test is a sufficient test with a selectable error. It can be regarded as a series of sufficient tests, which gets better with declining error.

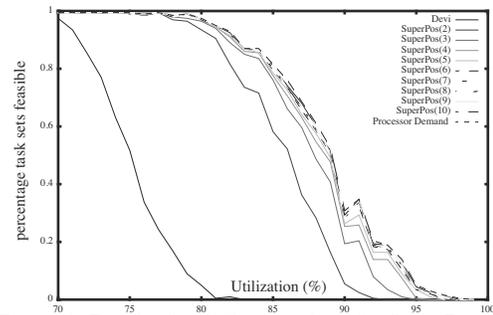

Figure 1: Results for different values for SuperPos(x)

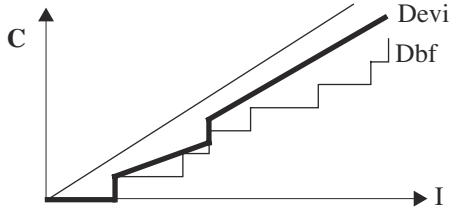

Figure 2: Approximation by Devi for two tasks

### 3.5. Superposition vs. Devi

An interesting point is the relationship of the superposition test to the best previously known sufficient test given by Devi [9].

**Lemma 2: Relationship Devi:** *A task set recognized as feasible by the test of Devi is also recognized as feasible by SuperPos(1).*

In other words, the test by Devi is only a special case of the superposition approach.

Proof: Consider the test *SuperPos(1)* (Lemma 1). Remember that in this special case the maximum test interval for each task is equal to their deadline ($I_m(\tau_i) = D_i$) as this is the ending time for the first job of the task. The test can be expressed as follow: ($\Gamma(i) = \{\tau_i \in \Gamma | (D_\tau \leq D_i)\}$

$$I \geq \sum_{\forall i | (D_i \leq I)} C_i + \sum_{\forall i | (D_i \leq I)} \frac{C_i}{T_i} \cdot (I - D_i)$$

$$= \sum_{\forall i | (D_i \leq I)} C_i + I \cdot \sum_{\forall i | (D_i \leq I)} \frac{C_i}{T_i} - \sum_{\forall i | (D_i \leq I)} \frac{D_i}{T_i} \cdot C_i$$

This test has to be preformed only for the first job of each task which has a deadline equal to the deadline of the task. Now consider the test of Devi (Definition 1). It can be expressed as: ($\forall (1 \leq k \leq |\Gamma|)$

$$D_k \geq D_k \cdot \sum_{i=1}^{k} \frac{C_i}{T_i} + \sum_{i=1}^{k} \left(\frac{T_i - min(T_i, D_i)}{T_i}\right) \cdot C_i$$

$$\geq D_k \cdot k \sum_{i=1}^{k} \frac{C_i}{T_i} + \sum_{i=1}^{k} \left(\frac{T_i - D_i}{T_i}\right) \cdot C_i$$

$$\geq \sum_{i=1}^{k} C_i + D_k \cdot \sum_{i=1}^{k} \frac{C_i}{T_i} - \sum_{i=1}^{k} \frac{D_i}{T_i} \cdot C_i$$

This test also has to be performed for the deadline of each task. Setting $I = D_k$ in the above equation results in *SuperPos(1)*. This proves Lemma 2. The relationship between the test by Devi and the superposition approach allows to include the extensions of the test by Devi described in [9] into the super position approach. The extensions concern practical relevant issues like switching time, priority ceiling protocol, self-suspension and limits for the number of priorities.

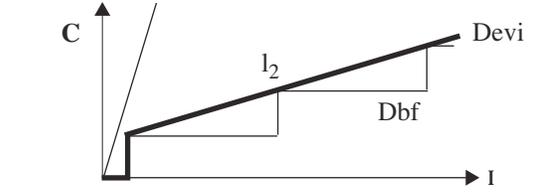

Figure 3: Approximation by Devi for a single task

### 3.6. Superposition vs. real-time calculus

In the concept of real-time calculus [7] based on the network calculus [6] the demand and the capacity of a system are described by arrival and service curves. The idea of these functions is the same as in the processor demand test, apart from that the capacity is not regarded as bisecting line but also as a function. The idea behind the real-time calculus is to define mathematical operations on the arrival and service curves to solve the feasibility test problem. By definition the curves are unlimited. Therfore the equations behind this concept are unfortunately expensive to compute in the case of general arrival and service curves [7]. To make the concept practicable a piecewise linear approximation with up to three staight line segments is proposed. The error of this approximation is unkown. By using the results of Section 3.5 it is possible to calculate a lower bound on the approximation error of the approximated real-time calculus. Figure 4 shows the real-time calculus approximation of one task. First the approximation of a simple periodic task with two lines is shown (a), second the approximation of a bursty task is outlined (b). In this case three lines are needed for a good approximation.

Comparing Figure 3 with Figure 4 a) shows a close relationship between the approximation of the real-time calculus and the test given by Devi. The real-time calculus approximation is a bit worse than the test given by Devi because of the limited number of curves in the approximation of the real-time calculus. As shown in the last section the test by Devi is equal to *SuperPos(1)* of the superposition approach. The main difference between the test by Devi and the real-time calculus is the specification of bursts which can be only expressed by the real-time calculus. By using event streams it is easy to analyse burst by the superposition test [1]. However, this leads to a higher complexity then the test by Devi because of each element of the burst has to be handled as a seperate element of the event stream.

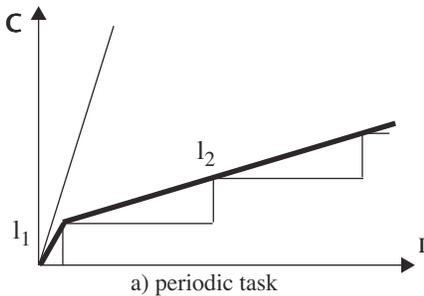

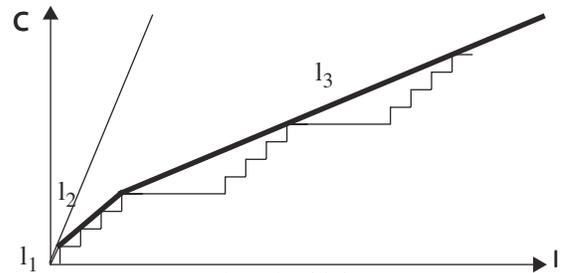

Figure 4: Real-time calculus approximation

## 4. Improved feasibility tests

Approximated feasibility tests are only sufficient. This is a problem because not all scheduable task sets are accepted by this tests. Even if the degree of sufficiency is selectable the tests fails to recognize feasible task sets (Figure 1). Even worse, choosing a high degree of sufficiency leads to a test with long execution time, while choosing a low degree leads to a bad acceptance rate for the feasible task sets. The idea is to use different levels of approximation, starting with a fast approximation (*SuperPos(1)*) and switch to a slow one just as necessary.

### 4.1. Dynamic error test

This idea leads to a fast test for all those task sets which the sufficient tests could recognize. The algorithm for the new test is shown in Figure 5. The test starts at the level *SuperPos(1)*. This level allows all tasks to be approximated after their first job. The test runs at this level until either the test succeeds or the approximated demand ($dbf'$) exceeds the actual test interval ($I_{act}$) and the task set is not feasible by this approximation. In this case it is necessary to rise the level and withdraw the approximations for those tasks which would not be approximated using the new level ($\Gamma_{rev}$). It is not necessary to recalculate the whole test because the values calculated by the first approximation can be reused.

**Lemma 3:** *If $dbf'(I,G) \leq I$ for an interval $I$ than $dbf(I',G) \leq I'$ for all intervals $I'$ between $I$ and $I_{next}$. $I_{next}$ is the next test interval after $I$ which is not approximated.*

The prove of this lemma is visualized in Figure 6 and can also be followed using Lemma 4. Therefore if the test fails at $I_{next}$, the previous test guarantees all intervals which are smaller than $I_{next}$, even if they are approximated.

```
ALGORITHM DynamicError
IF U > 1 ⇒ not feasible
Imax = minimum feasibility interval;
∀τ_i ∈ Γ :testlist.add(τ_t,T_i+ D_i)
WHILE (testlist ≠ {} AND I_act < Imax)
    τ_i, I_act = testlist.getNext()
    dbf' = dbf' + C_τ + (I_act - I_old) * Uready
    WHILE (dbf' > I_act)
        IF (ApproxList = {}) ⇒ NOT FEASIBLE
        increase level;
        ∀τ_j ∈ Γ_rev :  Uready = Uready - C_j / T_j,
                       dbf' = dbf' - app(I_act,τ_j)
                       testlist.add( τ_j, NextInt(I_act,τ_j))
    END WHILE
    IF (I_act < Testboarder(τ)
        testlist.add(τ, I_act + T_i)
    ELSE
        Uready = Uready + C_i / T_i
        ApproxList.add(τ);
    I_old =I_act
END WHILE
⇒ FEASIBLE
```

Figure 5: Dynamic Test

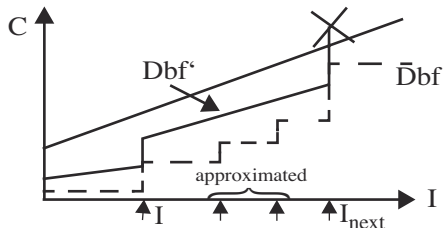

Figure 6: Possible proven test intervals

**Lemma 4:** *Let $I, I_{next}$ be two consecutive test intervals for $dbf'(I)$. If we assume it exists a $\Delta I$ for which $I < I + \Delta I < I_{next}$ and $dbf(I + \Delta I) > I + \Delta I$ applies then the approximated demand bound function holds $dbf'(I) > I$ [1].*

Proof: See Section 3.2 in [1].

For all tasks for which the approximation is revised due to the switch of the level the next test interval following $I_{act}$ has to be investigated:

**Lemma 5:** *Following test interval*

$$NextInt(I, \tau) = \left(\left\lfloor \frac{I - D_\tau}{T_\tau} \right\rfloor + 1\right) T_\tau + D_\tau$$

This interval is added to the list of test intervals. It is further necessary to reduce the approximated demand by the overestimated approximation costs for these revised tasks. These costs are given by *app*. Let $\Gamma_{rev}$ be the set of all tasks for which the approximation is revised:

**Lemma 6:** *Overestimated approximation costs*

$$app(I, \Gamma_{rev}) = \sum_{\forall \tau_i \in \Gamma_{rev}} \left( \frac{I - D_i}{T_i} - \left\lfloor \frac{I - D_i}{T_i} \right\rfloor \right) C_i$$

If $dbf' \leq I_{act}$ the new level is sufficient, otherwise it has to be increased. If no task is approximated before increasing the level the demand bound function exceeds the capacity and the test fails. We propose to double the level at each step which limits the amount of steps to $\sqrt{n_{max}}$. This is small compared to the total amount of test intervals.

Only task sets which cannot be evaluated using a fast level will need a long evaluation time. These task sets would not be accepted by existing sufficient tests like the test by Devi. Task sets accepted by the test by Devi are accepted by this test running completly on the level *SuperPos(1)*. This condition holds for all levels of the superpositon test. It is not necessary for a good average case performance of the analysis to combine sufficient and necessary tests. The maximum level for the dynamic test can be limited. The result is a test with a strictly limited worst-case run-time and a good average case run-time.

### 4.2. All approximated test

Despite that the proposed test outperform the existing approaches by orders of magnitude, it can be improved even further. Especially for task sets containing very small and very large tasks, the dynamic test could switch soon into higher levels. Considering many test intervals may result in an unnecessary effort. It could be the case that only a few test intervals are critical and that it is possible to approximate the others. This leads to a new test. The algorithm for this test can be found in Figure 7.

The goal of the new algorithm is to reduce the number of considered test intervals further. Figure 7 shows the algorithm. Instead of using fixed test bounds for each task, approximation is done as much as possible. The first test intervall resulting out of the first job of each task is inserted into *testlist*. *Testlist* is processed in ascending order. All the following test intervalls are approximated first. Only if a test fails for an test interval $I_{test}$, the approximation of the *demand bound function* of each task is step by step revised until the test either succeds or no task is approximated any more. The revision is done by replacing the approximated costs of the task by their real cost. Because of quitting the approximation in this interval it is necessary to add one

```
ALGORITHM AllApprox
IF U > 1 ⇒ not feasible
∀τ_i ∈ Γ : testlist.add(τ,T_i + D_i)
WHILE (testlist ≠ {})
    τ_i, I_test = testlist.getNext()
    dbf' = dbf' + C_τ + (I_test - I_old) * Uready
    WHILE (dbf' > I_act)
        IF (ApproxList = {}) ⇒ not feasible
        τ' = ApproxList->getAndRemoveFirstTask;
        Uready = Uready - C_τ' / T_τ';
        dbf' = dbf' - app(I_test,τ')
        testlist.add(τ', NextInt(I_test,τ'))
    END WHILE
    Uready = Uready + C_i / T_i
    ApproxList.add(τ);
END WHILE
⇒ feasible
```

Figure 7: All approximated test

additional test interval for each revised task into *testlist*. These additional test intervals can be calculated using Lemma . Note that it is only necessary to calcultate the interval following $I_{test}$ because the approximation has already verified all test intervals smaller than $I_{test}$. The algorithm terminates if it is possible to approximate all tasks successfull in one test interval or if the test fails. If the initial test interval is accepted for each task without generating new test intervals, the behaviour and the performance of the test is equal to the test given by Devi [9].

### 4.3. Feasibility bound

A feasibility bound describes the upper limit for all intervals which are necessary to be tested. It is important to find a short one because this limits the effort for the feasibility test. A good overview of the different feasibility bounds can be found in [14]. Baruah et al. [3] defined such a bound for the processor demand test, which is part of Definition 3. Another one can be derived from the busy period condition [14]. George et al. [10] define a bound which is smaller than the bound given by Baruah et al:

$$I_{george} = \frac{\sum_{\tau_i \in \Gamma \wedge D_i \leq T_i} \left(1 - \frac{D_i}{T_i}\right) \cdot C_i}{1 - U}$$

The new all approximated superpositon approach delivers a new feasibility bound. It is reached (and no further test interval is needed) if for a test interval the difference between the demand bound function and the capacity allows the approximation of all tasks. A task system with a utilization lower 100% can never exceed the capacity beyond this interval. Using Lemma 6 the sum of the demand bound function and the approximation errors can be calculated:

$$I \geq \sum_{\tau \in \Gamma} app(I, \tau) + dbf(I, \tau)$$

This gives the new feasibility bound:

$$I_{sup} = \sum_{\tau \in \Gamma} \left(\frac{I_{sup} - D_\tau}{T_\tau} - \left\lfloor \frac{I - D_\tau}{T_\tau} \right\rfloor\right) C_\tau + \sum_{\substack{\forall \tau_i \in \Gamma \\ I \geq D_\tau}} \left\lfloor \frac{I - D_\tau}{T_\tau} + 1 \right\rfloor \cdot C_\tau$$

If the condition $I_{sup} \geq D_{max}$ is used, it can be followed:

$$I_{sup} = I_{sup} \cdot \sum_{\tau \in \Gamma} \frac{C_\tau}{T_\tau} - \sum_{\tau \in \Gamma} \frac{C_\tau \cdot D_\tau}{T_\tau} + \sum_{\tau \in \Gamma} C_\tau$$

Using $U = \sum C/T$ leads to the following bound:

$$I_{sup} = min(D_{max}, \frac{\sum_{\forall \tau_i \in \Gamma} (1 - D_i/T_i) \cdot C_i}{1 - U})$$

For the case that all $C_\tau \leq D_\tau$ this bound is the same as the bound given by George et al., for the other cases this new bound is lower than this bound. This relationship allows a better understanding of the bound and leads to a integrated feasibility theory for EDF. The main contribution of this result is that it is not necessary for the new test to calculate and check the feasibility bound by George et al. The superposition bound is tighter and is checked in an implicit way. It might be interesting to calculate the busy period of the system as outlined in [14], because this bound might be tighter than the superposition bound under some conditions. Calculating this bound however has exponential complexity and may need more effort than the test without this bound.

## 5. Experiments

We have generated a large number of random task sets to evaluate the performance gain due to the new tests. As a metric we counted the test intervals checked by each of the algorithms. This metric is common to evaluate feasibility tests [5], [13] and shows exactly the behavior of the different algorithms.

The generation of the random task sets follows the uniform distribution proposed by Bini [4]. In Figure 8 a test was performed using 18,000 task sets with a utilization between 90% and 99% (high utilizations are hard to test). It shows the average and the maximum iterations needed for the tests. The size of the task sets varied between 5 and 100 tasks using a uniformed random distribution. They had an average gap of 20%, 30% and 40%. The gap describes the difference between deadline and period. The sizes of periods are also equally distributed, the ratio between the maximum and the minimum period was of no concern for this experiment. The result of the experiment is that the all approximation test needs 10 to 20 times less iterations than the processor demand test, which goes up to 200 times less effort considering the maximum amount of iterations.

Figure 9 shows the result of another experiment which investigates the effect of the ratio $T_{max}/T_{min}$. Therefore the test was performed with a ratio varying from 100 to 1,000,000. The results are presented using an exponential scale. Such high ratios could for example be possible if system interrupts and the schedulability overhead are defininded as tasks. For each value 4,000 tasks set where generated randomly with a size between 5 and 100 tasks and an average gap between 10% and 50% and an utilization varying between 90% and 100%. The first graph shows the maximum measured effort needed for the processor demand test which increases up to more than 50 million iterations. The effort of the other values is visualized in the second graph. Also there are some variations due to the random values especially for the dynamic test. The measured effort of both new tests are far below the effort needed for the old test. The maximum effort measured for the dynamic test was about 9,000. For the all approximation test it is about 3,000 iterations. Another interesting point is that the effort doesn't depend on the ratio of the periods. This becomes even more obvious when investigating the average effort needed. For the all approximation test it varies between 102 and 116 iterations, as opposed to 465 to 955,053 for the processor demand test.

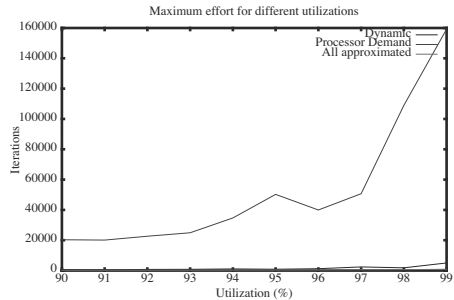
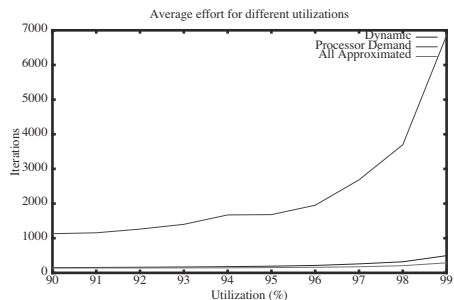

Figure 8: Results with different utilizations

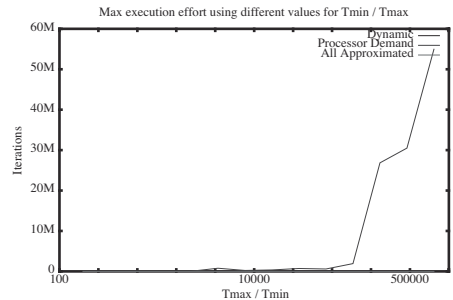
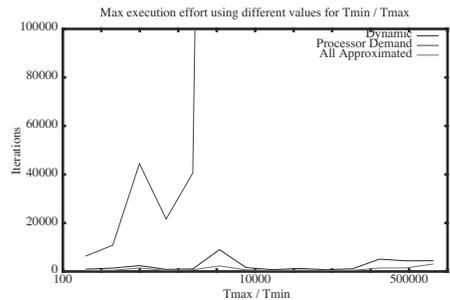

Figure 9: Results with different values for $T_{min}$ / $T_{max}$

Tab. 1 contains some results for task sets coming from real examples. The task set of Burns and of the modified set of Ma & Shin can be found in [1], the *Generic Avionics Platform* example (GAP) in [14] and the other two task sets in [11]. The amount of tasks are small (7 to 21 tasks) but even for such small examples the new tests need between 5 and 100 times less iterations than the processor demand test. The run-time overhead of one iteration ot the new tests is small compared to both alternative algorithms, the processor demand test and the test by Devi. Only the approximation leads to a some more effort. Compared to the algorithm of Devi the overhead can be completely eliminated.

| Test | Devi | Dyn. | All Appr. | Proc. Dem. |
|---|---|---|---|---|
| Burns | 14 | 14 | 14 | 1,112 |
| Ma & Shin | FAILED | 16 | 11 | 61 |
| GAP | 18 | 18 | 18 | 1,228 |
| Gresser 1 | FAILED | 24 | 20 | 307 |
| Gresser 2 | FAILED | 34 | 25 | 205 |

Table 1: Iterations for example task graphs

## 6. Conclusion

We proposed two new feasibility test for *EDF*. They are exact and outperform the existing tests by orders of magnitude. Task sets which are recognized by the existing sufficient test or approximations can be evaluated by the new test with a comparable effort. We proved that the new concept integrates the processor demand test, the feasibility bound by George et al. and the sufficient feasibility test by Devi.